\documentclass[preprint,12pt]{elsarticle}
\usepackage{setspace}

 \usepackage{graphicx}

\usepackage{amssymb}
 \usepackage{amsthm}

\journal{arXiv.org quant-ph}

\begin{document}

\begin{frontmatter}

\title{Ququat as superposition of coherent state and their application in quantum information processing}

\author{Manoj K Mishra$^1$, Hari Prakash$^2$ and Vibhuti Bhushan Jha$^1$}

\address{$^1$Space Applications Center,Indian Space Research Organisation, Ahmedabad, India\\$^2$Physics Department, University of Allahabad, Allahabad, India\\
$^2$Indian Institute of Information Technology, Allahabad, India\\
Email: manoj.qit@gmail.com}

\begin{abstract}

Superposition of optical coherent states $\left|\pm\alpha\right\rangle$, possessing opposite phases, play an important role as qubits (quantum state defined in a two dimensional Hilbert space) in quantum information processing tasks like quantum computation, teleportation, key distribution etc., and are of fundamental importance in testing quantum mechanics. Passage of such superposition of coherent states from a 50:50 beam splitter lead to generation of entangled coherent states. Recently, ququats and qutrits defined in four and three dimensional Hilbert space respectively, have attracted much attention as they offer advantage in secure quantum communication. However, practical utilization of these advantages essentially require physical realization of quantum optical ququats and qutrits. 

Here, we define four new multi-photonic states with $4n +j$ (here, $j =0, 1, 2$ or $3$, and $n = 0, 1, ..., \infty$) numbers of photon and show how the superposition of coherent states can be used to encode ququat using these multi-photonic states as basis vectors of a four dimensional Hilbert space. When these multi-photonic states fall upon a 50:50 beam splitter, the resulting states are bipartite four-component entangled coherent states equivalently representing the entangled ququats. We briefly discuss the photon statistical properties of such multi-photonic states and bipartite four-component entangled coherent states. We show that these multi-photonic states and bipartite four-component entangled coherent states can be synthesized using even coherent states as input to an interferometer. We give a simple linear optical protocol for almost perfect teleportation of a ququat encoded in superposition of coherent states with the aid of bipartite four-component entangled coherent state as quantum channel. We also describe how these ququats can be used for realization of higher dimensional BB84 protocol in order to increase the security of quantum key distribution. Finally, we discuss the possible advantages of using superposition of coherent states as ququats and bipartite four-component entangled coherent states as quantum channel in different quantum information processing tasks.

\end{abstract}


\end{frontmatter}


\section{Introduction}
\label{}
Peculiar characteristics of quantum mechanics like the superposition principle, quantum entanglement, no-cloning theorem, no-deleting theorem and the uncertainty principle led to the development of pure quantum information processing protocols like quantum teleportation (QT) ~\cite{1}, quantum key distribution ~\cite{2}, quantum superdense coding ~\cite{3} and quantum secret sharing ~\cite{4}. The possibility of QT ~\cite{1} (i.e., transferring an unknown quantum state of one system to another system across space using quantum entanglement ~\cite{5}) is of utmost interest in linear quantum optical computation ~\cite{6, 7} and in secure quantum communication ~\cite{8, 9}. Different kind of optical systems have been studied as candidates for encoding quantum information and for quantum information processing. For example, in a laboratory experiment by Bouwmeester \emph{et al.} ~\cite{10}, standard bi-photonic entangled states (the Bell states ~\cite{11}) were used to realize the QT of single photon qubit. The success rate of QT in this experiment was equal to $0.5$ due to the fact that only linear optical elements were used for the Bell state measurement. In another experiment by Kim \emph{et al.} ~\cite{12}, unit success rate was reported for QT of single photon qubit with the aid of nonlinear sum frequency generation interaction for the Bell state measurement. On the other hand, Boschi \emph{et al.} ~\cite{8}, transformed the standard bi-photonic entangled state to a k-vector entangled state by passing each entangled photon through a calcite crystal and then k-vector entangled state was used for QT of single photon qubit. Recent achievements in teleporting single photon qubit over large distance using fiber optical links ~\cite{14,15} and free space links ~\cite{16, 17} promises for future satellite based quantum communication. However, the most difficult task in commercial realization of QT using standard bi-photonic entangled states as quantum channel is an efficient realization of the Bell state measurement, in which four standard bi-photonic entangled states should be discriminated. It was shown that the four Bell states of the form of standard bi-photonic entangled states cannot be discriminated using linear optical elements ~\cite{18,19}, which makes it difficult to achieve a unit success rate. Knill \emph{et al.} ~\cite{7} proposed a realization of universal gate operations based on linear optics and photon detection, but the requirement of nonlinear interactions to perform complete Bell state measurement is one of the major hindrances to the implementation of deterministic gate operations as well as scalable quantum computer.  

In recent past another form of entangled state, entangled coherent state (ECS) ~\cite{20, 21, 22}, has attracted much attention. Gerry ~\cite{20} and Sanders ~\cite{21} proposed nonlinear Mach-Zehnder interferometer as a device to transform a pair of coherent state to ECS of the form of $\left|E\right\rangle =N_{\pm}\left(\left|\alpha,\alpha\right\rangle \pm \left|-\alpha,-\alpha\right\rangle\right)$ and $\left|E\right\rangle =N_{\pm}\left(\left|\alpha,-\alpha\right\rangle \pm \left|\alpha,-\alpha\right\rangle\right)$ where, $N_{\pm}$ refer to appropriate normalization constants and $\left|\alpha\right\rangle$  refers to the coherent state of radiation field. Howell and Yeazell ~\cite{22} have proposed generation of ECSs via two non-demodulation measurements. Munro et al ~\cite{23} suggested a scheme for encoding entangled qubit as ECS of two-dimensional center of mass vibrational motion for two trapped ions. Proposal for entanglement transfer from the two vibrational modes to the electronic states of the two ions in order for the Bell state to be detected by resonance fluorescence shelving method has also been suggested ~\cite{23}. Entanglement properties of ECSs has been studied ~\cite{24, 25, 26, 27}. Recently few studies have discussed the dynamics of quantum discord in ECSs ~\cite{28} and quasi-Werner states based on ECSs ~\cite{29}. It was shown that ECSs are much stronger against decoherence due to photon absorption than the standard bi-photonic entangled states~\cite{27}. It has been explicitly shown that all four ECSs can be well discriminated using only linear optical elements (a beam splitter and two photon number resolving detectors), which is not the case with standard bi-photonic entangled states. This has become a remarkable advantage for using ECSs as quantum channel for quantum teleportation, quantum key distribution, quantum superdense coding and quantum computation. Various quantum computation schemes using coherent state as qubit ~\cite{30, 31, 32} including deterministic gate operations with ECSs as offline resources ~\cite{30} have been suggested. Macroscopically distinct even and odd superposition of coherent states (SCS) given by $\left|\pm\right\rangle =n_{\pm}\left(\left|\alpha \right\rangle \pm \left|-\alpha\right\rangle\right)$ can be used as logical qubit encodings for quantum computation ~\cite{33}. In even and odd SCS based encoding scheme, a single decay event due to amplitude damping transforms an even (odd) SCS to odd (even) SCS  which appears as a bit flip error. Thus by selecting appropriate amplitude damping time, error correction can be performed~\cite{33}. On the other hand, a single decay event on the Fock superposition state leads to a state that cannot be recovered by unitary operation. This offers advantage of using even and odd SCS based encoding scheme not only for quantum computation, but also for other quantum information processing tasks. Another advantage of using SCS as qubit is that they circumvent the necessity of perfect single photon sources. Enk and Hirota ~\cite{34} have proposed scheme for teleportation of a SCS using ECS with success probability equal to 0.5. Wang ~\cite{35} has presented a very similar scheme for teleporting bipartite ECS with success probability equal to 0.5. Furthermore, Prakash \emph{et al.} ~\cite{36} modified the photon counting scheme and reported almost perfect teleportation for an appreciable mean photon number. The effect of de-coherence on fidelity and quality of teleportation has also been studied in ~\cite{37}. Schemes for Quantum teleportation of four-component bipartite ECS have been suggested in ~\cite{38, 39, 40}. Qauntum teleportation of SCS using non-maximally ECSs  and the effect of entanglement on teleportation fidelity with different unitary operation strategies to be adopted by the receiver to recover replica of information state with as large a fidelity as possible has also been proposed~\cite{41}. Various proposals have been suggested for quantum metrology ~\cite{42, 43, 44, 45} and a quantum repeater ~\cite{46} with SCS and ECS. The utility of SCS of radiation field with more than two components has been investigated for sensitive force detection ~\cite{45}. Quantum key distribution schemes based on coherent state ~\cite{47, 48}, SCS ~\cite{49} and ECS ~\cite{50} have been proposed by various researchers. A very good review on quantum information processing with SCS and ECS can be found in ~\cite{51}. It is to be noted that the most basic requirement for the physical realization of above mentioned schemes is the availability of SCS and ECS of appropriate coherent amplitude. In this direction number of schemes have been proposed for generation of SCS [~\cite{52}-~\cite{56}] which in turn can be used for deterministic generation of ECSs just by using an additional symmetric beam splitter.

In most of these works, qubits have been used as basic element for encoding information and entangled qubits have been used as quantum channel to accomplish different quantum information processing tasks. However, higher dimensional quantum systems (qudits and entangled qudits) have attracted much interest in recent past for their larger capacity of information encoding and transmission as compared to qubits and entangled qubits. It has been shown in various studies that higher dimensional systems (e.g., ququats and qutrits) present advantage in secure quantum communication and in researches on foundation of quantum mechanics [~\cite{57}-~\cite{60}]. Being defined in a two dimensional Hilbert space, polarization is not suitable for the implementation of qudits and entangled qudits. Few schemes have been proposed for realization of qudits and entangled qudits in optics based on other degree of freedom of photons, for example, orbital angular momentum entangled qutrits ~\cite{61,62} and polarization of two-photon entangled qudits [~\cite{63}-~\cite{69}]. Orbital angular momentum ~\cite{70} is an interesting candidate for encoding qudits since it is defined in an infinite dimensional Hilbert space. However, the orbital angular momentum of light is still lacking of a complete set of tools which allow to manipulate and detect it as simply as it is done with polarization. Qing ~\cite{71} has presented an optical scheme to generate entangled qutrits by coding the state in propagation path of photons. The success probability for generation of such maximally entangled qutrits and ququats was equal to 0.6 and 0.5, respectively. Thus the scheme proposed by Qing ~\cite{71} was probabilistic in nature with low success rate. The shortcoming of using such entangled qutrits or ququats is that these cannot be discriminated using linear optical elements only, which was the case with standard bi-photonic entangled states. For these reasons the superdense coding scheme ~\cite{71} based on photon propagation path entangled qutrit or ququat is probabilistic in nature, leading to a decreased information transmission capacity (log3 bits using entangled qutrits). Moreover a single decay event will transform the photon propagation path entangled qutrit or ququat to a state that cannot be recovered by unitary operation, thus error correction becomes difficult. Also photon path propagation based qutrits and ququats are not the promising candidates to realize higher dimensional BB84 protocol in order to increase the security. Scheme for generation of four component SCS of vibrational modes of a trapped ion was suggested in Ref. ~\cite {72}, however, scheme for encoding ququat in such SCS has not been proposed. For these reasons, it is required to investigate new candidates for encoding quantum information and quantum entanglement in higher dimensional system such that: a) they can be deterministically generated, b) can be used for various quantum information applications and  c)discrimination of the entangled states should require linear optical elements only. To date, single mode SCS have been employed only for encoding qubit using even and odd SCS as basis states of two dimensional Hilbert space. However, fundamental advantages of using higher dimensional quantum states for quantum information encoding, transmission and processing and for testing quantum mechanics necessiates the need to study SCS as a possible candidate. Motivated by this, in the present paper we study the possibility of encoding one ququat in a single mode SCS, generation of entangled ququats based on coherent states and protocols for QT and quantum key distribution based on SCS encoded with one ququat.

The rest of the paper is organized as follows. In Section (2), we first define four new multi-photonic states (MPS) $\left|\alpha_{j}\right\rangle$ with $4n +j$ (here, $j =0, 1, 2$ or $3$, and $n = 0, 1, ..., \infty$) number of photons and show how these superposition of coherent states $\left|\pm\alpha\right\rangle$ and $\left|\pm{i}\alpha\right\rangle$,  which are $90^{o}$ out of phase can be used as basis vectors to project an infinite dimentional Hilbert space to a four dimensional Hilbert space. We also define four bipartite four-component entangled coherent states (BFECS) which are another kind of ECSs and show that these BFECS equivalently represent non-maximally entangled ququats in $4\otimes4$ dimensional Hilbert space spanned by MPS as basis states.  We briefly discuss the photon statistical properties of MPS and BFECS. In Section (3), we present an optical scheme to generate such MPS and BFECS.  In section (4) we show how a ququat can be  encoded in SCS using MPS as basis vectors of a four dimensional Hilbert space and also show how BFECS can be used as useful resource for quantum information processing by constructing a linear optical protocol for faithful teleportation of single ququat encoded in superposition of optical coherent states. In section (5), we estimate over all quality of our QT scheme. In section (6), we show how MPS may be used for realization of higher dimensional BB84 protocol to enhance the security of quantum key distribution.  In section (7), we address issues related to our ququat based QT and quantum key distribution scheme. We also discuss the possible advantage of using BFECS for superdense coding over photon propagation path based entangled ququats. Finally in section (8), we present the conclusions.

\begin{figure}[ht!]
\centering
\includegraphics[width=13cm]{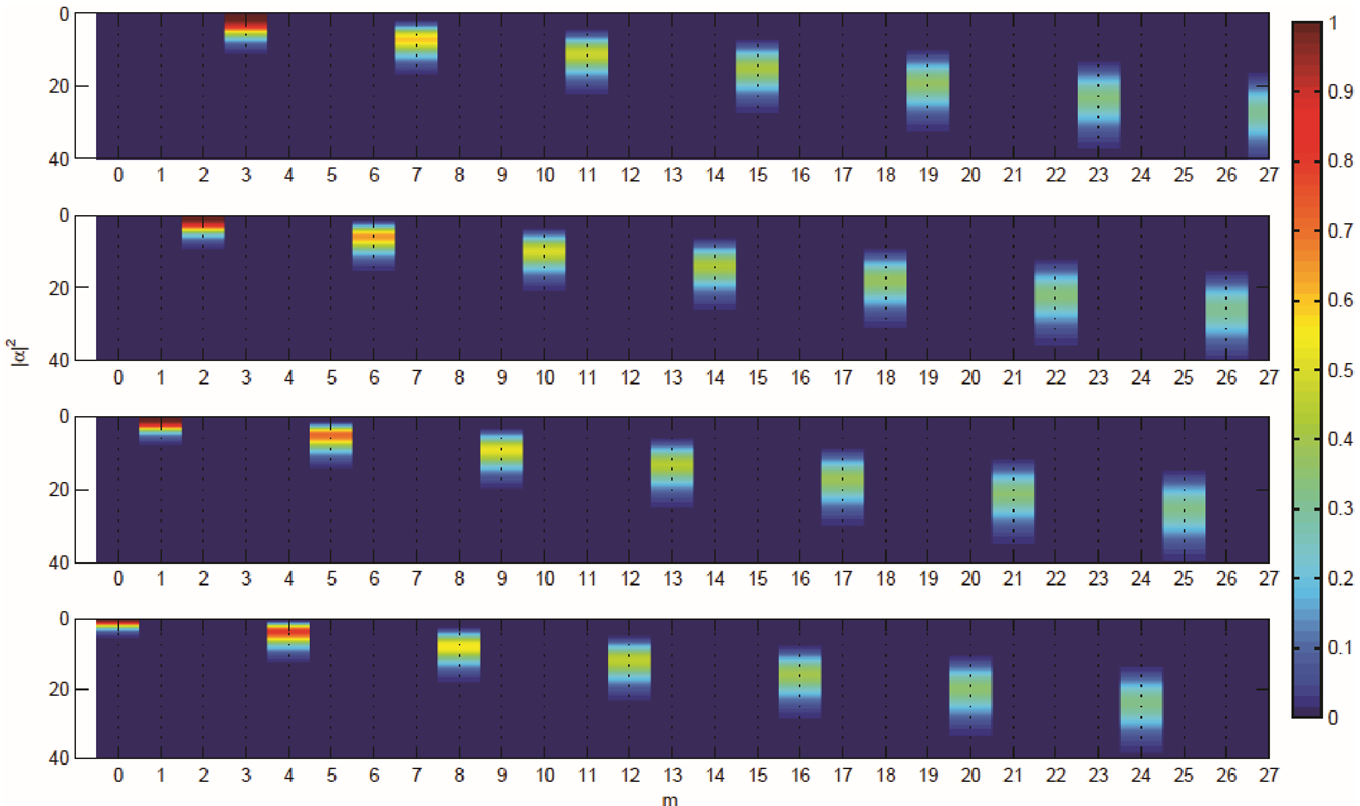}
\caption{Color map for probability of occurrence of $m$ number of photons in multi-photonic states $\left|\alpha_{0}\right\rangle$, $\left|\alpha_{1}\right\rangle$, $\left|\alpha_{2}\right\rangle$ and $\left|\alpha_{3}\right\rangle$ (in bottom to top order) with respect to mean photon numbers}
\label{Fig.1}
\end{figure}

\section{Photon statistical properties of MPS and BFECS}
\label{}
Let us first define four orthonormal MPS states as
\begin{equation}
\label{eq1}
\left| {\alpha _j } \right\rangle =N_j \sum\nolimits_{m=0}^3{(-i)^{jm}\left| {i^m\alpha } \right\rangle } ,\,\,j=0,1,2,3,
\end{equation}
with
\begin{equation}
\label{eq2}
\begin{array}{l}
N_{0,2} =[2(1+x^2\pm 2x\cos \left| \alpha \right|^2)^{1/2}]^{-1},\\
N_{1,3}=[2(1-x^2\pm 2x\sin \left| \alpha \right|^2)^{1/2}]^{-1},
\end{array}
\end{equation}
where $x=\exp (-\left| \alpha \right|^2)$ and $\left| \alpha \right\rangle =\exp (-\left| \alpha \right|^2)\sum\nolimits_{n=0}^\infty {(\alpha ^n/\sqrt {n\,!} } )\left| n \right\rangle $ is the usual Glauber coherent state of radiation with coherent amplitude $\left| \alpha \right|$ ~\cite{73}. In this whole paper, index $j$ runs from 0 to 3. The probability of finding $m$ number of photons in MPS $\left|\alpha_{j}\right\rangle$ is given by $P(\alpha_{j},m)=\left|\left\langle m|\alpha_{j}\right\rangle\right|^{2}=4\left|N_{j}
\right|^{2}x\left|\alpha\right|^{2}(m!)^{-1}[1+(-1)^{m+j}][1+\sqrt{2}cos\left\{(2j+1)\pi/4\right\}cos\left\{(m+mod(j,1))\pi/2\right\}]$. Fig.~\ref{Fig.1} shows $P(\alpha_{j},m)$ for different values of photon number $m$ (x-axis) and for different values of mean photon number of the order of $\left|\alpha\right|^{2}$ (y-axis). It is evident from Fig.~\ref{Fig.1} that MPS $\left| {\alpha _j} \right\rangle $ gives non-zero probability only for 4$n +j$ number of photons, where $n$ = 0,1,...,$\infty$. Using eq.~(\ref{eq1}), we can write
\begin{equation}
\label{eq3}
\left| {i^j\alpha } \right\rangle =\textstyle{1 \over 
2}\sum\nolimits_{m=0}^3 {(i)^{jm}r_m \left| {\alpha _m } \right\rangle } ,
\end{equation}
where $r_{j}$ = 1/(2$N_{j})$. Therefore any coherent state defined in an infinite dimensional Hilbert space spanned by photon number states, can equivalently be defined in a 4 dimentional Hilbert space spanned by states $\left|{\alpha _j } \right\rangle $. ECS of the form $\left|ECS_{0,1}\right\rangle\approx(\left|\alpha,\alpha\right\rangle\pm\left|-\alpha,-\alpha\right\rangle)$
and $\left|ECS_{2,3}\right\rangle\approx(\left|\alpha,-\alpha\right\rangle\pm\left|-\alpha,\alpha\right\rangle)$ have been extensively studied by many authors ~\cite{24, 25, 26, 27} and have been shown to be useful resource for quantum 
information processing ~\cite{30}-~\cite{40}. However, here we define four different type of ECS called bipartite four-component entangled coherent states (BFECS) as
\begin{equation}
\label{eq4}
\left| {E_j } \right\rangle =N_{E _{j}} \sum\nolimits_{m=0}^3 {(-i)^{jm}\left| {i^m\alpha ,i^m\alpha } \right\rangle } ,
\end{equation}
where 
\begin{equation}
\label{eq5}
\begin{array}{l}
N_{E _{0,2}} =[2(1+x^4\pm 2x^2\cos 2\left|\alpha \right|^2)^{1/2}]^{-1},\\
N_{E _{1,3}} =[2(1-x^4\pm 2x^2\sin 2\left|\alpha \right|^2)^{1/2}]^{-1}.
\end{array}
\end{equation}
The BFECS in eq.~(\ref{eq4}) can be expanded in photon number state basis as $\left| {E_j } \right\rangle =4xN_{E _{j}}\sum_{n,m|n+m=4k+j}[\alpha^{n+m}/\sqrt{n!m!}]\left|n,m\right\rangle$ where $k=0,1,2,...,\infty$.  One can verify that the difference between ECS and the BFECS is that in former the total number of photons is always even or odd, whereas in the later it is always $4k+j$. We also note that by expanding the BFECS in  eq.~(\ref{eq4}) in MPS basis using eq. (\ref{eq3}), BFECS can be written in the form of non-maximally entangled ququat as 
\begin{equation}
\label{eq6}
\left| {E_j } \right\rangle =N_{E _{j}} \sum\nolimits_{m=0}^3 {r_m r_{j-m+4} 
\left| {\alpha _m ,\alpha _{j-m+4} } \right\rangle },
\end{equation}
where $j-m+4$ = mod( $j-m+4$, 3). For coherent amplitudes in the limit $\left| \alpha \right|\to \infty $, $N_{Ej}$ and $r_{j}$ reduce to 0.5 and 1.0, respectively. Remarkably, for $\left| \alpha \right|\to \infty$ BFECS become maximally entangled ququat as it can be easily verified that the reduced density matrix obtained by taking trace over any one mode is maximally mixed with four nonzero eigenvalues, all equal to 1/4. This shows that BFECS for appreciable coherent amplitude hav approximately 2-ebit of entanglement. 

\section{Generating entangled ququat or BFECS}
\label{}
\begin{figure}[ht!]
\centering
\includegraphics[width=6cm]{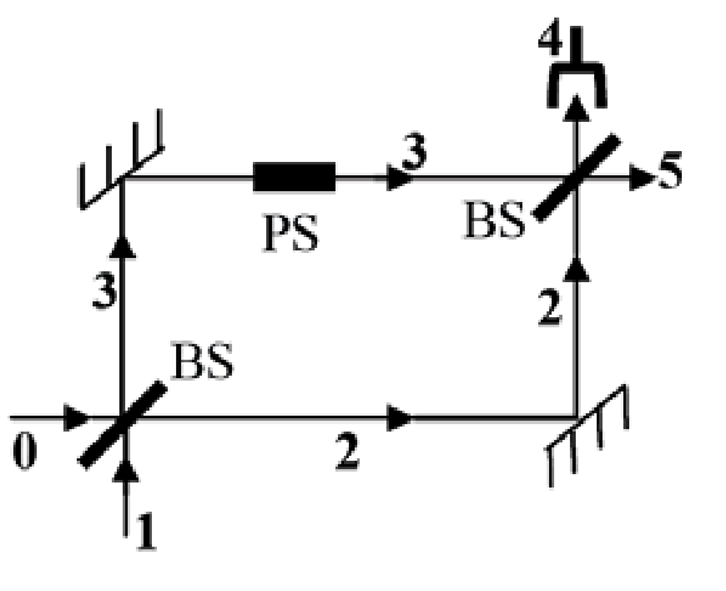}
\caption{Scheme for generating states $\left| {\alpha _j } \right\rangle $ where $j$ = 0, 1, 2, 3 (eq.~(\ref{eq1})) having 4$n+j$ numbers of photon. BS and PS stands for symmetric beam splitter and -$\pi $/2 phase shifter, respectively. Bold numbers represent the quantum mode. Here, even superposition of coherent states $\left| + \right\rangle _0 =N_e [\left| \alpha \right\rangle +\left| {-\alpha } \right\rangle ]_0 $ and $\left| {{+}'} \right\rangle _1 =N_e [\left| {-i\alpha } \right\rangle +\left| {i\alpha } \right\rangle ]_1 $ are used as inputs to interferometer in modes 0 and 1, respectively.}
\label{Fig.2}
\end{figure}

The BFECS in eq.~(\ref{eq4}) can be produced from MPS in eq.~(\ref{eq1}) by splitting it in a lossless symmetric beam splitter. When a symmetric beam splitter is illuminated by two coherent states 
$\left| \alpha \right\rangle _{in1} \left| \beta \right\rangle _{in2} $ in modes \textit{in}1 and\textit{in}2, output state in modes \textit{out}1 and \textit{out}2 is given by $\left|{2^{-1/2}(\alpha +i\beta )} 
\right\rangle _{out1} \left| {2^{-1/2}(i\alpha +\beta )} \right\rangle _{out2} $. A -$\pi $/2 phase shifter converts state $\left| \alpha \right\rangle $ to $\left| {-i\alpha } \right\rangle $. Consider two even superposition of coherent states $\left| + \right\rangle _0 =N_e [\left| \alpha \right\rangle +\left| {-\alpha } \right\rangle ]_0 $ and $\left| {{+}'} \right\rangle _1 =N_e [\left| {-i\alpha } \right\rangle +\left| {i\alpha } \right\rangle ]_1 $ as inputs to an interferometer in modes 0 and 1, respectively, as shown in fig.~\ref{Fig.2}. Here normalization constant is given by $N_{e}=$[2(1+$x^{2})$]$^{-1/2}$. The output state is given by
\begin{equation}
\label{eq7}
\begin{array}{l}
\left| \psi \right\rangle _{4,5} =N_e ^2\sum\nolimits_{m=0}^3 {\left| {i^m\alpha ,(-i)^{m+3}\alpha } \right\rangle } _{4,5} =N_e^2\sum\nolimits_{m=0}^3 {i^mr_m ^2\left| {\alpha _m ,\alpha _m } 
\right\rangle } _{4,5} .
\end{array}
\end{equation}
Eq.~(\ref{eq7}) shows that photon counting in mode 4 gives 4$n +j$ number of photon counts where $j $ may take values $0, 1, 2 or 3$, corresponding to which state $\left| {\alpha _j } \right\rangle $ gets generated in mode 5 with probability $P_{j }=N_{e}^{4}r_{j}^{4}$. $P_{j}$ becomes equal to 0.25 for appreciable value of $\left| \alpha \right|$. After illuminating a symmetric beam splitter by state $\left| {\alpha _j } \right\rangle $, the resulting state is an entangled ququat similar to BFECS $\left| {E_j } \right\rangle $, with coherent amplitudes equal to $\left| \alpha \right|/\sqrt 2 $. 

\begin{figure*}[ht!]
\centering
\includegraphics[width=12.5cm]{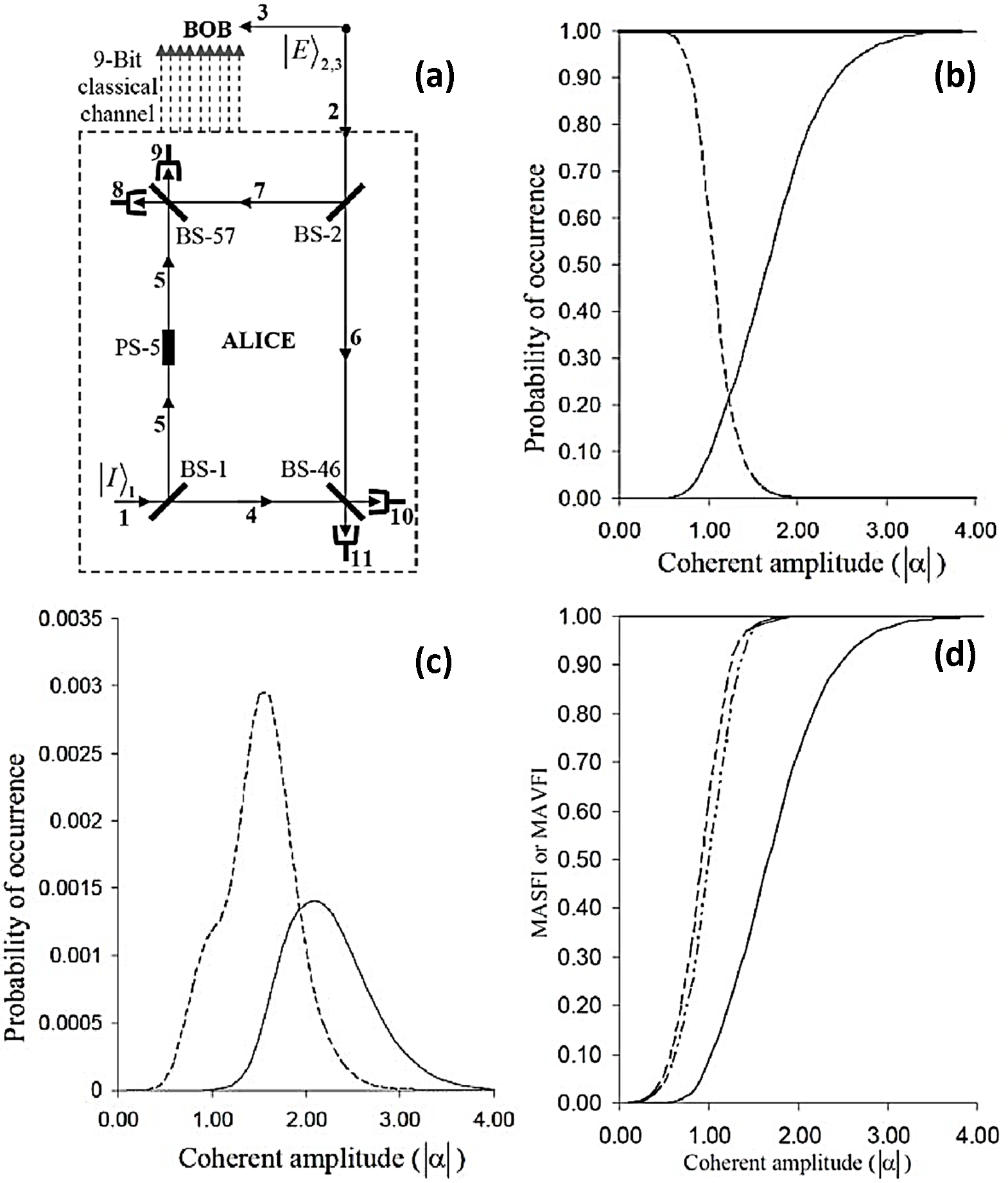}
\caption{(a) Scheme for teleporting one ququat encoded in SCS (eq.~(\ref{eq8} or \ref{eq9})) using entangled ququat called BFECS (eq.~(\ref{eq4})). BS and PS stand for symmetric beam splitter and -$\pi $/2 phase shifter, respectively. Bold numbers represent the quantum mode. (b) Dashed and continuous curve show maximum 
value of $P_{I}$ and the summation of probabilities of occurrence for all 256 PC results belonging to Group IV, respectively. (c) Dashed and continuous curves shows maximum value of probabilities $P_{II}$ and $P_{III.I}^{ }$, respectively. (d) Dashed, dash-dotted and continuous curves shows minimum 
assured fidelities (MASFI) $F_{5,6,8,10}^{MASFI} $,$F_9^{MASFI} $ and minimum average fidelity (MAVFI), respectively.}
\label{Fig.3}
\end{figure*}
\section{Encoding ququat in SCS and their teleportation}
\label{}
The information state to be teleported from Alice to Bob is given by
\begin{equation}
\label{eq8}
\left| I \right\rangle _1 =\sum\nolimits_{m=0}^3 {\varepsilon _m \left| {i^m\alpha } \right\rangle } _1 ,
\end{equation}
with $\sum\nolimits_{m=0}^3[\left|{\varepsilon_{m}}\right|^2+x^{2}\varepsilon_{m}^{\ast}\varepsilon _{\underline{m+2}}+x(\varepsilon^{\ast}_{m}\varepsilon _{\underline{m+1}}x^{-i}+\varepsilon _{m}^{\ast} \varepsilon _{\underline{m+1}}x^{i})] =1$ as normalization condition, 
where $\varepsilon _m $ are the complex coefficients and ($m$+1)=mod($m$+1,3), ($m$+2)=mod($m$+2,3). Using eq.~(\ref{eq3}) information state $\left| I \right\rangle _1$ can be expanded in MPS basis as 
\begin{equation}
\label{eq9}
\left| I \right\rangle _1 =\sum\nolimits_{m=0}^3 {c_m \left| {\alpha _m } \right\rangle } _1 ,
\quad
\sum\nolimits_{m=0}^3 {\left| {c_m } \right|^2=1} ,
\end{equation}
where $c_m =r_m \sum\nolimits_{k=0}^3 {i^{mk}\varepsilon _k } /2$. Clearly, eq.~(\ref{eq9}) represents an arbitrary ququat.

Any of the BFECS in eq.~(\ref{eq4}) can in principle be used for QT a ququat. The QT protocol is shown in fig.~\ref{Fig.3}(a). The protocol shown in fig.~\ref{Fig.3}(a) is chosen because of its simplicity; it is to be noted that it is not an optimistic scheme. We shall see it does however lead to almost perfect QT of a ququat for appreciable coherent amplitude. The information mode 1 is with Alice. Entangled modes 2 and 3 of BFECS $\left| {E_0 } \right\rangle _{2,3} $ (eq.~(\ref{eq4})) are with Alice and Bob, respectively. The initial state of the system is
\begin{equation}
\label{eq10}
\begin{array}{l}
\left| \psi \right\rangle _{1,2,3} =\left| I \right\rangle _1 \otimes \left| {E_0 } \right\rangle _{2,3}
=N_{E_0 } \sum\nolimits_{m,n=0}^3 {\varepsilon_m \{\left| {i^m\alpha ,i^n\alpha ,i^n\alpha } \right\rangle \}}.
\end{array}
\end{equation}
Effect of the QT protocol on an arbitrary initial state $\left|{a,b,c} \right\rangle _{1,2,3} $ is described by
\begin{equation}
\label{eq11}
\begin{array}{l}
\left| {a,b,c} \right\rangle _{1,2,3} \to \left|{\frac{i(a+b)}{2},\frac{(a-b)}{2},\frac{(a+ib)}{2},\frac{(ia+b)}{2},c} \right\rangle _{8,9,10,11,3}
\end{array}
\end{equation}
Using this for each component on right hand side of eq.~(\ref{eq10}), the final output state is given by
\begin{equation}
\label{eq12}
\begin{array}{l}
\left| \psi \right\rangle _{8,9,10,11,3} =N_{E_{0}} \sum\nolimits_{m=0}^3 {\varepsilon _m }
 [\left| {0,i^m\alpha ,i^{m+3}\beta ,i^{m+1}\beta ,i^{m+2}\alpha } \right\rangle\\
 +\left| {i^{m+1}\alpha ,0,i^m\beta ,i^m\beta ,i^m\alpha } \right\rangle  
 +\left| {i^{m+1}\beta ,i^{m+3}\beta ,0,i^{m+1}\alpha ,i^{m+1}\alpha } \right\rangle \\
 +\left| {i^m\beta ,i^m\beta ,i^m\alpha ,0,i^{m+3}\alpha }\right\rangle ]_{8,9,10,11,3} \\ 
\end{array}
\end{equation}
where $\left| \beta \right\rangle =\left| {(1+i)\alpha /2} \right\rangle $. The output modes 8, 9, 10 and 11 are with Alice and mode 3 is with Bob. Alice performs photon counting in modes 8, 9, 10 and 11 and conveys her photon counting result to Bob, on the basis of which Bob performs an appropriate unitary operation in mode 3 to retrieve replica of the information state. 

Coherent states are the superposition of all photon number states; therefore, there will be many photon counting results. For better understanding of all photon counting results, we expand coherent states $\left| {i^j\alpha } \right\rangle $ with
Alice into states $\left| 0 \right\rangle $, $\left| {\alpha _1 } \right\rangle $, $\left| {\alpha _2 } \right\rangle $, $\left| {\alpha _3 } \right\rangle $, and $\left| {\alpha _4 } \right\rangle $ with 0, 
4$n$+1,4$n$+2,4$n$+3, and 4$n$+4 ($n = 0,1, ..., \infty$ ) numbers of photon, respectively, as
\begin{equation}
\label{eq13}
\left| {i^j\alpha } \right\rangle =x^{1/2}\left| 0 \right\rangle +\sum\nolimits_{m=1}^4 {i^{jm}a_m } \left| {\alpha _m } \right\rangle ,
\end{equation}
where $a_k =r_k /2\,\,(k=1,2,3)$, $a_4 =\textstyle{1 \over 2}(r_0^2-4x)^{1/2}$. States $\left| {\alpha _{k=1,2,3} } \right\rangle $ are given in eq.~(\ref{eq1}), while $\left| {\alpha _4 } \right\rangle $ is given by
\begin{equation}
\label{eq14}
\left| {\alpha _4 } \right\rangle =N_4 [\left| \alpha \right\rangle +\left|{i\alpha } \right\rangle +\left| {-\alpha } \right\rangle +\left| {-i\alpha} \right\rangle -4\sqrt x \left| 0 \right\rangle ],
\end{equation}
where 
\begin{equation}
\label{eq15}
N_4 =[2(1-4x+x^2+2x\cos \left| \alpha \right|^2)^{1/2}]^{-1}.
\end{equation}
Similarly coherent states, $\left| {i^j\beta } \right\rangle $ can be expanded into states $\left| 0 \right\rangle $, $\left| {\beta _1 } \right\rangle $, $\left| {\beta _2 } \right\rangle $, $\left| {\beta _3 } 
\right\rangle $, and $\left| {\beta _4 } \right\rangle $ with 0, 4$l$+1, 4$l$+2, 4$l$+3, and 4$l$+4 (here $l$ = 0, 1, ..., $\infty )$ numbers of photon, respectively, as
\begin{equation}
\label{eq16}
\left| {i^j\beta } \right\rangle =x^{1/4}\left| 0 \right\rangle +\sum\nolimits_{m=1}^4 {i^{jm}b_m } \left| {\beta _m } \right\rangle .
\end{equation}
States $\left| {\beta _{m=1,2,3,4} } \right\rangle $ and coefficients $b_{m=1,2,3,4} $ can be defined by substituting $\left| \alpha \right|^2/2$ instead of $\left| \alpha \right|^2$ in expressions for states $\left|{\alpha _{m=1,2,3,4} } \right\rangle $ and coefficients $a_m $. 

eq.~(\ref{eq12}) shows that one of the modes 8, 9, 10 and 11 always give zero count as photon counting result. eqs.~(\ref{eq12}), (\ref{eq13}) and (\ref{eq16}) tell that each of the other three modes can give any of the five results, zero or nonzero which is 0, 1, 2 or 3 modulo 4. Thus, there are $^{4}$C$_{1}$4$^{3 }+^{4}$C$_{2}$4$^{2 }$+ $^{4}$C$_{3}$4$^{1 }+^{4}$C$_{4}$4$^{0 }$= 369, different photon counting results. These results can be transmitted to Bob through a 9-bit classical channel. We write these photon counting results as 0, 1, 2, 3 and 4, the last one being the nonzero- result (0 modulo 4) written as 4 to distinguish it from the result 
of 0 counts.

These results can be classified into four groups: Group I (\textit{All modes count zero photon}), Group II (\textit{Any three modes count zero and one mode count non-zero photon}), Group III (\textit{Any two modes count zero photon and rest two modes count non-zero photon}), and Group IV (\textit{Only one mode count zero and rest three modes count non-zero photons}). Fidelity of the teleported state is a measure 
of quality of QT and it is defined as the overlap of teleported state ($\left| T \right\rangle )$ over the original information state ($\left| I \right\rangle ) \quad F=\left| {\left\langle {T} \mathrel{\left| {\vphantom {T I}} \right. \kern-\nulldelimiterspace} {I} \right\rangle } \right|^2$. Since information state is unknown, therefore it is necessary to introduce the minimum assured fidelity which is defined as minimum of the fidelity over all possible information states. Putting
\begin{equation}
\label{eq17}
\begin{array}{l}
c_0 =\cos \theta \cos \phi _1 ,
c_1 =\cos \theta \sin \phi _1 e^{i\xi _1 },
\quad\\
c_2 =\sin \theta \cos \phi _2 e^{i\xi _2 }),
c_3 =\sin \theta \sin \phi _2 e^{i\xi _3 },
\end{array}
\end{equation}
the minimum assured fidelity is obtained by minimizing $F$ over $\theta ,\,\phi _1 ,\,\phi _2 ,\,\xi _1 ,\,\xi _2 ,\xi _3 $. The overall quality of QT scheme is measured by average fidelity which is defined as summation of the product of fidelity and probability of occurrence of all possible photon counting results, $F_{av} =\sum\nolimits_k {P_k F_k } $.

\textbf{Group I:} This group has only one photon counting result and after photon counting, the state with Bob in mode 3 is $\left| {B_0 } \right\rangle =\left| {\alpha _0 } \right\rangle $. No unitary operation can be prescribed to retrieve original information, therefore QT fails. The probability of occurrence for this photon counting result is given by 
\begin{equation}
\label{eq18}
P_I =(4x^2\left| {c_0 } \right|^2)/(1+x^4+2x^2\cos 2\left| \alpha 
\right|^2).
\end{equation}
Fig.~\ref{Fig.3}(b) shows that the maximum value of $P_{I}$ becomes almost zero for $\left| \alpha \right|\ge 1.5$, therefore this failure will not affect the average fidelity for $\left| \alpha \right|\ge 1.5$.

\textbf{Group II:} This group has $^{4}$C$_{1}$4 = 16 possible photon counting results as the non-zero photon mode may be any one of the four modes and non-zero photon counts may be any of 4$n$+1, 4$n$+2, 4$n$+3 or 4$n$+4 (here $n$ = 0, 1, ..., $\infty)$. Considering one of these, viz, when mode 8 gives 4$n$+4 numbers of photon count and rest mode give zero counts, we found that it is not possible to predict the state with Bob after photon counting and prescribe a unitary transformation. Therefore, we admit failure. The probability of occurrence is given by
\begin{equation}
\label{eq19}
\begin{array}{l}
P_{II} =N_{E_0 }^2 [a_4 ^2x+4b_4 ^2x^{3/2}(\left| {c_0 } \right|^2+\left| {c_2 } \right|^2)\\
+4x^2\{\cos (\left| \alpha \right|^2/2)\cosh (\left| \alpha \right|^2/2)-1\}(\left| {c_0 } \right|^2-\left| {c_2 } \right|^2)].
\end{array}
\end{equation}
Fig.~\ref{Fig.3}(c) shows that the maximum value of $P_{II}$ become almost zero for $\left| \alpha \right|\ge 2.8$. Similar results are obtained for the rest 15 photon counting results in this Group. Hence failure of this Group will not degrade the average fidelity for $\left| \alpha \right|\ge 2.8$.

\textbf{Group III:} This group has $^{4}$C$_{2}$4$^{2}$ = 96 photon counting results, which are further divided into two subgroups, Subgroup III.I and Subgroup III.II. 

\emph{Subgroup III.I:} (Pair of modes '8 and 10' or '8 and 11' or '9 and 10' or '9 and 11' show zero counts, while the rest two modes show non-zero photons): This subgroup has $^{4}$C$_{1}$4$^{2}$ = 64 photon counting results. Considering one of these that gives 4$n$+4 counts in modes 8 and 10, while rest zero, we found that it is not possible to predict the state with Bob in mode 3 after photon counting. Therefore, teleportation fails. However, the probability of occurrence for this photon counting result is given by
\begin{equation}
\label{eq20}
\begin{array}{l}
P_{III.I} =2N_{E_0 }^2 [a_4 ^2b_4 ^2x^{1/2}+x^2\{\cos (\left| \alpha \right|^2/2)
\cosh (\left| \alpha \right|^2/2)-1\}^2(\left| {c_0 } \right|^2-\left| {c_2 } \right|^2)].
\end{array}
\end{equation}
Fig.~\ref{Fig.3}(c) shows that the maximum value of $P_{III.I}$ become almost zero for $\left| \alpha \right|\ge 3.2$. Similar conclusions are obtained for the rest 63 of photon counting results. Thus failure of this subgroup will not affect the average fidelity for $\left| \alpha \right|\ge 3.2$.

\emph{Subgroup III.II:} (modes '8 and 9' or '10 and 11' counts zero, while rest mode count non-zero photons): This subgroup has $^{2}$C$_{1}$4$^{2}$ = 32 photon counting results. If we look at the states with Bob for the 32 photon counting results, it is seen that the Bob's state is invariably in the form
\begin{equation}
\label{eq21}
B^{(j,k,m)}=(1/2)[B^{(j,k)}+i^mB^{(j+2,k)}],
\end{equation}
where 
\begin{equation}
\label{eq22}
B^{(j,k)}=\sum\nolimits_{l=0}^3 {c_{l+k} } (r_l /r_{l+k} )i^{jl}\left| {\alpha _l } \right\rangle .
\end{equation}
For the cases where a unitary transformation resulting in perfect or almost perfect teleportation exists, we write the required unitary transformations for the Bob's state $B^{(j,k,m)}$ as 
\begin{equation}
\label{eq23}
U^{(j,k,m)}=(1/2)[U^{(j,k)}+(-i)^mU^{(j+2,k)}],
\end{equation}
where 
\begin{equation}
\label{eq24}
U^{(j,k)}=\sum\nolimits_{l=0}^3 {(-i)^{jl}\left| {\alpha _{k+l} } 
\right\rangle } \left\langle {\alpha _l } \right|.
\end{equation}
Here, indices $j$, $k$, $m$ = 0, 1, 2 or 3. For the cases where no unitary transformation giving $F $= 1 is possible and minimum assured fidelity equals to zero, we admit failure, but prescribe unitary transformations $U^{(j,k)}$ which give $F $= 1 for certain cases of special information states, although minimum assured fidelity equals to zero. For more details see Appendix A, where all 32 photon counting results, corresponding Bob's states, unitary transformations and fidelities are tabulated in table A.1.

The teleported states are seen to occur in one of the six different forms given as
\begin{equation}
\label{eq25}
\begin{array}{l}
\left| {T_1 } \right\rangle =c_0 \left| {\alpha _0 } \right\rangle +c_2 \left| {\alpha _2 } \right\rangle ,\left| {T_2 } \right\rangle =c_1 \left|{\alpha _1 } \right\rangle +c_3 \left| {\alpha _3 } \right\rangle , \\ 
 \left| {T_3 } \right\rangle =c_3 r_1 /r_3 \left| {\alpha _3 } \right\rangle +c_1 r_3 /r_1 \left| {\alpha _1 } \right\rangle ,\\
\left| {T_4 } \right\rangle =c_2 r_0 /r_2 \left| {\alpha _2 } \right\rangle +c_0 r_2 /r_0 \left| {\alpha 
_0 } \right\rangle, \\ 
 \left| {T_5 } \right\rangle =\sum\nolimits_{m=0}^3 {c_m r_{m+3} /r_m } \left| {\alpha _m } \right\rangle ,\\
\left| {T_6 } \right\rangle =\sum\nolimits_{m=0}^3 {c_m r_{m+1} /r_m } \left| {\alpha _m } \right\rangle 
\\ 
\end{array}
\end{equation}
where $m$+3=mod($m$+3,3) and $m$+1=mod($m$+1,3). The corresponding fidelities $F_{1,2,3,4,5,6} $ are given by
\begin{equation}
\label{eq26}
\begin{array}{l}
 F_1 =\left| {c_0 } \right|^2+\left| {c_2 } \right|^2,\,\,F_2 =\left| {c_1 } \right|^2+\left| {c_3 } \right|^2,\\
F_3 =[\left| {c_3 } \right|^2r_1 ^2+\left| {c_1 } \right|^2r_3 ^2]^2[\left| {c_3 } \right|^2r_1 ^4+\left| 
{c_1 } \right|^2r_3 ^4]^{-1}, \\ 
 F_4 =[\left| {c_2 } \right|^2r_0 ^2+\left| {c_0 } \right|^2r_2 ^2]^2[\left| {c_2 } \right|^2r_0 ^4+\left| {c_0 } \right|^2r_2 ^4]^{-1},\\
F_5 =[\sum\nolimits_{m=0}^3 {\left| {c_m } \right|} ^2r_{m+3}^2 r_{m+2} r_{m+1}]
[\sum\nolimits_{m=0}^3 {\left| {c_m } \right|} ^2r_{m+3}^4 r_{m+2}^2 
r_{m+1}^2 ]^{-1}, \\ 
 F_6 =[\sum\nolimits_{m=0}^3 {\left| {c_m } \right|} ^2r_{m+1}^2 r_{m+2}r_{m+3}]
 [\sum\nolimits_{m=0}^3 {\left| {c_m } \right|} ^2r_{m+1}^4 r_{m+2}^2 r_{m+3}^2 ]^{-1}. \\ 
 \end{array}
\end{equation}
For $F_{1,2,3,4}$ the minimum assured fidelity vanishes i.e., $F_{1,2,3,4}^{MASFI} =0$, while fig.~\ref{Fig.3}(d) shows that $F_{5,6}^{MASFI} $ reaches unity for $\left| \alpha \right|\ge 1.7$. Thus for 16 photon counting results QT fails, while rest 16 photon counting results give almost perfect QT for $\left| \alpha \right|\ge 1.7$. 

\textbf{Group IV:} This group has $^{4}$C$_{1}$4$^{3}$ = 256 photon counting results. For this group of photon counting results the Bob's state and unitary transformation are seen to occur in the form $B^{(j,k)}$ and $U^{(j,k)}$, respectively, defined earlier by eqs.~(\ref{eq22}) and (\ref{eq24}). For more details see Appendix A, where all 256 photon counting results, corresponding Bob's state and required unitary transformation, are tabulated in table A.2 and A.3. Defining 
\begin{equation}
\label{eq27}
\begin{array}{l}
\left| {T_7 } \right\rangle =\sum\nolimits_{m=0}^3 {c_m } \left| {\alpha _m } \right\rangle ,
\left| {T_8 } \right\rangle =\left| {T_5 } \right\rangle ,\\
\left| {T_9 } \right\rangle =\sum\nolimits_{m=0}^3 {c_m r_{m+2} /r_m } \left| {\alpha _m } \right\rangle ,
\left| {T_{10} } \right\rangle =\left| {T_6 } \right\rangle ,
\end{array}
\end{equation}
where $m$+2=mod($m$+2, 3). It is found that there are 64 cases each for $k$ = 0, 1, 2 and 3. For all photon counting results belonging to $k$ = 0, teleported state is $\left| {T_7 } \right\rangle $with fidelity $F_{7}$ =1, therefore QT is perfect. For photon counting results belonging to $k$ = 1, 2 and 3, teleported states are $\left| {T_8 } 
\right\rangle $, $\left| {T_9 } \right\rangle $ and $\left| {T_{10} } \right\rangle $ with fidelities $F_8 $, $F_9 $ and $F_{10} $, respectively, given as
\begin{equation}
\label{eq28}
\begin{array}{l}
F_8 =F_5 ,\,\,F_{10} =F_6 ,\\
F_9 =[\sum\nolimits_{m=0}^3 {\left| {c_m } \right|} ^2r_{m+2}^2 r_{m+1} r_{m+3} ]
[\sum\nolimits_{m=0}^3 {\left| {c_m } 
\right|} ^2r_{m+1}^4 r_{m+2}^2 r_{m+3}^2 ]^{-1}.
\end{array}
\end{equation}
Fig.~\ref{Fig.3}(d) shows that the minimum assured fidelity $F_{8,9,10}^{MASFI} $ becomes almost equal to unity for $\left| \alpha \right|\ge 1.7$. Thus 64 photon counting results give perfect QT for any value of $\left| \alpha \right|$, while rest 192 photon counting results gives almost perfect QT for $\left| \alpha \right|\ge 1.7$. 
\section{Overall quality of QT}
\label{}
Fig.~\ref{Fig.3}(b) shows that the summation of probability of occurrence of all 256 photon counting results belonging to Group IV becomes almost equal to unity for $\left| \alpha \right|\ge 3.2$. Therefore for $\left| \alpha \right|\ge 3.2$, any photon counting result will essentially be one among 256 photon counting results belonging to Group IV, which gives perfect or almost perfect QT. Thus the protocol described here guarantees at least almost perfect QT with 
almost perfect success rate for $\left| \alpha \right|\ge 3.2$. Average fidelity is calculated and minimized over all possible information state to give minimum average fidelity. Fig.~\ref{Fig.3}(d) manifests that minimum average fidelity of QT becomes almost unity, i.e., minimum average fidelity becomes $\ge $ 0.99 for $\left| \alpha \right|\ge 3.2$.

\section{Quantum key distribution using MPS}
Quantum key distribution first proposed by Bennett and Brassard in 1984 (BB84 protocol) provides a way to distribute a secret key between two distant legitimate users Alice (the sender) and Bob (the receiver), with no assumptions of computational power of an eavesdropper, Eve ~\cite{2}. Here, we show that higher dimensional BB84 protocol can be implemented with MPS states (ququats based on SCS) defined in section (2). Let us define two mutually unbiased bases B1 and B2, corresponding to 8 state vectors. The first base B1 is composed of four orthogonal MPS states (see eq. ~(\ref{eq1}))
\begin{equation}
\label{eq29}
\begin{array}{l}
B1\equiv\left\{\left|\alpha_{0}\right\rangle, \left|\alpha_{1}\right\rangle, \left|\alpha_{2}\right\rangle, \left|\alpha_{3}\right\rangle\right\}.
\end{array}
\end{equation}
The second base B2 is composed of superposition of MPS states
\begin{equation}
\label{eq30}
B2\equiv\left\{
\begin{array}{l}
\frac{1}{2}\left(\left|\alpha_{0}\right\rangle+\left|\alpha_{1}\right\rangle+\left|\alpha_{2}\right\rangle+\left|\alpha_{3}\right\rangle\right),\\    \frac{1}{2}\left(\left|\alpha_{0}\right\rangle+e^{\frac{\pi}{2}i}\left|\alpha_{1}\right\rangle+e^{{\pi}i}\left|\alpha_{2}\right\rangle+e^{-\frac{\pi}{2}i}\left|\alpha_{3}\right\rangle\right),\\
\frac{1}{2}\left(\left|\alpha_{0}\right\rangle+e^{{\pi}i}\left|\alpha_{1}\right\rangle+\left|\alpha_{2}\right\rangle+e^{{\pi}i}\left|\alpha_{3}\right\rangle\right), \\ \frac{1}{2}\left(\left|\alpha_{0}\right\rangle+e^{-\frac{\pi}{2}i}\left|\alpha_{1}\right\rangle+\left|\alpha_{2}\right\rangle+e^{\frac{\pi}{2}i}\left|\alpha_{3}\right\rangle\right).
\end{array}
\right\}
\end{equation}
In section (2) we described a scheme that enables us to generate any state vector belonging to base B1. One can verify that for appreciable value of coherent amplitude $\left|\alpha\right|$ the normalization constants $N_{0,2}$ and $N_{1,3}$ defined in eq.~(\ref{eq2}) becomes almost equal to 0.5. For this reason, when eq. ~(\ref{eq1}) is used to represent the state vectors of base B2 in terms of coherent states, the base B2 reduces to
\begin{equation}
\label{eq31}
\begin{array}{l}
B2\equiv\left\{\left|\alpha\right\rangle, \left|i\alpha\right\rangle, \left|-\alpha\right\rangle, \left|-i\alpha\right\rangle\right\}
\end{array}
\end{equation}
for appriciable value of coherent amplitude $\left|\alpha\right|$. Thus in the limit of large $\left|\alpha\right|$ four state vectors belonging to base $B2$ can be generated simply by applying appropriate phase shifter on a laser pulse represented by coherent state $\left|\alpha\right\rangle$. To start key distribution Alice encodes her "quat" (1 quat equals to 2 bits) information by choosing randomly among eight states vectors belonging to two bases B1 and B2, and transmits them to Bob. Now Bob randomly chooses the base B1 or B2 in which he wants to measure the encoded "quat". If Bob's choice is B1, he can do so just by photon counting. On the other hand if Bob's choice is B2, he first splits the state vector in to two parts by using a 50/50 symmetric beam splitter. Assuming that the transmitted state by Alice is $\left|\gamma\right\rangle$ the output state in mode 1 and 2 is given by $\left|\gamma/\sqrt{2}\right\rangle_{1}\left|\gamma/\sqrt{2}\right\rangle_{2}$. In second step, Bob mixes mode 1 and 2 with two ancillary coherent states given by $\left|\alpha/\sqrt{2}\right\rangle_{1}\left|i\alpha/\sqrt{2}\right\rangle_{2}$ in mode 3 and 4, respectively. Doing this the final output state is given by
\begin{equation}
\label{eq32}
\left|\gamma\right\rangle\rightarrow\left|\left(\gamma+\alpha\right)/2\right\rangle_{5}\left|\left(\gamma-\alpha\right)/2\right\rangle_{6}\left|\left(\gamma+i\alpha\right)/2\right\rangle_{7}\left|\left(\gamma-i\alpha\right)/2\right\rangle_{8}
\end{equation}
It can be verified that if $\left|\gamma\right\rangle$ belongs to base B2, the final output states using these transformations are given by
\begin{equation}
\label{eq33}
\begin{array}{l}
\left|\gamma\right\rangle=\left|\alpha\right\rangle\rightarrow\left|\alpha\right\rangle_{5}\left|0\right\rangle_{6}\left|\left(1+i\right)\alpha/2\right\rangle_{7}\left|\left(1-i\right)\alpha/2\right\rangle_{8}\\
\left|\gamma\right\rangle=\left|i\alpha\right\rangle\rightarrow\left|\left(1+i\right)\alpha/2\right\rangle_{5}\left|\left(-1+i\right)\alpha/2\right\rangle_{6}\left|i\alpha\right\rangle_{7}\left|0\right\rangle_{8}\\
\left|\gamma\right\rangle=\left|-\alpha\right\rangle\rightarrow\left|0\right\rangle_{5}\left|-\alpha\right\rangle_{6}\left|\left(-1+i\right)\alpha/2\right\rangle_{7}\left|\left(-1-i\right)\alpha/2\right\rangle_{8}\\
\left|\gamma\right\rangle=\left|-i\alpha\right\rangle\rightarrow\left|\left(1+i\right)\alpha/2\right\rangle_{5}\left|\left(-1+i\right)\alpha/2\right\rangle_{6}\left|0\right\rangle_{7}\left|-i\alpha\right\rangle_{8}.
\end{array}
\end{equation}
From above given equation it is clear that one mode amongst 5, 6, 7 and 8 always gives vacuum state, therefore four states belonging to base B2 can be discriminated by performing photon counting in modes 5, 6, 7 and 8. Thus Bob can perform measurement in both bases B1 and B2 just by using few linear optical elements and photon counting. Without Eve, if n ququats are sent, on an average Bob guesses the right base in half of the cases, so that n/2 shared "quats" (n bits) are perfectly correlated. On the other hand, if Eve uses a standard intercept-resend strategy to gain information about the shared key, it is easy to show that on an average she gets half of the transmitted bits, just like in the analogous protocol in two dimensions. But since Eve's choice of measurement bases is also random, on an average Eve guesses the wrong base in half of the cases, therefore induceing disturbance in the communication leading to an error rate in Bob's results. Here ququats are used for encoding information; therefore each wrong guess about measurement base by Eve will unavoidably introduce an error rate equal to 3/4. Now since probability of guessing wrong base is 1/2, therefore, total error rate induced by Eve in Bob's results is equal to 3/8, which is larger than in the qubit case (3/8 against 1/4).

\section{Discussion}
It is to be noted that QT protocol given in section (4)  is not an optimistic one. The reason behind this is the kind of measurement scheme we have adopted. Since one mode on Alice side is always in vacuum state eq.~(\ref{eq12}) and since vacuum cannot be perfectly distinguished from coherent state especially at low coherent amplitude, therefore we considered states $0, \left|\alpha_1\right\rangle, \left|\alpha_2\right\rangle, \left|\alpha_3\right\rangle$ and $\left|\alpha_4\right\rangle$ as measurement basis on Alice side instead of considering MPS in eq.~(\ref{eq1}) as our measurement basis. Another reason why we adopted such measurement basis is that it can be achieved simply by photon counting which makes our QT protocol simple but not optimistic. However it is straight forward to formulate the optimum QT protocol using MPS in eq.~(\ref{eq1}) as measurement basis but it would require measurements that are harder to implement. It is to be noted that similar to the case of two dimension, where Pauli gate (X, Y and Z gates) operations on qubits can be realized by QT of qubit using entangled qubits, here also from table (2) we can see that present QT protocol can be used for realization of four dimensional Pauli gates.

In section (6) we have shown that MPS can be used to realize a higher dimensional BB84 protocol to increase the security of distributed key as compared to that achievable with a two dimensional BB84 protocol. In our scheme photon numbers (base B1) and relative phases (base B2) have been used to encode a 'quat'. Since the photon number and phase are conjugate variables, therefore any attempt by Eve to select two photon pulses (an essential step to perform photon number splitting attack ~\cite{74,75}) will unavoidably randomize the phase and photon counting result of the pulse and hence information encoded in pulse phase gets destroyed before Eve can make any measurement to extract information. In each such attempt Eve introduces the error to the Bob's measurement result, while in case of intercept resend attack, among all eavesdropping attempts, on average Eve introduces error only in half of the attempts. For these reasons, our scheme is robust against photon splitting attack and we considered intercept resend attack for security analysis. Since vacuum cannot be perfectly distinguished from coherent states, therefore whenever Bob decides to measure the received ququat in base B1 there is a nonzero probability of getting inconclusive result. Therefore Bob and Alice have to discard such inconclusive events leading to a decrease in key distribution rate of our scheme. However, since the probability of having vacuum in a coherent state decreases to almost zero for coherent amplitude greater than equal to 2, therefore for appreciable coherent amplitude inconclusive events will not affect the key distribution rate. 

As discussed earlier in introduction section that the quantum superdense coding scheme with photon propagation path entangled qutrits and ququats becomes probabilistic in nature due to the fact that all such entangled states can not be discriminated using linear optical elements. For these reasons the information transmission capacity reduces to log3 and log4=2 bits with the case of photon propagation path entangled qutrits and ququats, respectively. On the other hand from four BFECS $\left| {E_j }\right\rangle_{1,2} $ defined in eq.~(\ref{eq4}), we can construct four more BFECS $\left| {E^{'}_{j}}\right\rangle_{1,2}$ by applying a 180 degree phase shifter $P(\pi)$ in mode 1. $P(\pi)$ coverts a coherent state $\left|\alpha \right\rangle$ to $\left|-\alpha \right\rangle$. Thus BFECS $\left| {E^{'}_{j}}\right\rangle_{1,2}$ are given by 
\begin{equation}
\label{eq34}
\left| {E^{'}_{j}} \right\rangle_{1,2} =P(\pi)\left| {E_j } \right\rangle_{1,2}=N_{E _{j}} \sum\nolimits_{m=0}^3 {(-i)^{jm}\left| {i^m\alpha ,-i^m\alpha } \right\rangle_{1,2} }.
\end{equation}
The advantage of using these eight BFECS ($\left| {E_{j}} \right\rangle_{1,2}$ and $\left| {E^{'}_{j}} \right\rangle_{1,2}$) for quantum information processing tasks is that these can be completely discriminated simply by mixing the two entangled modes over a 50/50 symmetric beam splitter followed by photon counting in output modes. It can be verified that when entangled modes 1 and 2 of eight BFECS are mixed over a 50/50 symmetric beam splitter, the output states in mode 3 and 4 are given by
\begin{equation}
\label{eq35}
\begin{array}{l}
\left| {E_{j}} \right\rangle_{1,2}\rightarrow\left| \alpha^{'}_{j},0 \right\rangle_{3,4}\\
\left| {E^{'}_{j}} \right\rangle_{1,2}\rightarrow\left| 0,\alpha^{'}_{j} \right\rangle_{3,4}
\end{array}
\end{equation}
where, states $\left| \alpha^{'}_{j},0 \right\rangle$ are the MPS $\left| \alpha_{j},0 \right\rangle$ (see eq.~(\ref{eq1})) with different coherent amplitude. Eq.~(\ref{eq35}) shows that photon counting results in mode 3 and 4 are orthogonal to each other and makes it clear that all eight BFECS can be discriminated using linear optical elements. Thus, one can perform superdense coding protocol using BFECS with information transmission capacity equal to 3 bits which is higher than 2 bits achievable by using photon propagation path entangled ququats. Inability to discriminate photon propagation path entangled ququats using linear optical elements, will also avoids us to achieve teleportation success rate equal to unity with such entangled states, on the other hand, our QT scheme with the aid of BFECS gives almost unit success using linear optical elements only.

To complete the discussion it is necessary to address the issue of decoherence. The effect of decoherence due to photon absorption (amplitude damping) will decrease the entanglement of BFECS. In such case the teleported state will get entangled to the environment and hence the fidelity of teleportation will decrease. However, this is not the case with BFECS quantum channels only. Any entangled state have to suffer the effect of decoherence. However, BFECS are more robust against decoherence due to photon absorption as compared to single photon based entangled states (for example, photon propagation path entangled ququats) because one can verify by using the master equation approach ~\cite{33} that a single photon decay event will lead to sudden death of entanglement in single photon based entangled state, while BFECS entanglement decreases gradually with decay time. It can also be verified by using the solution of master equation for amplitude damping given in ~\cite{33} that a single decay event will convert the MPS states $\left|\alpha_{0}\right\rangle\rightarrow \left|\alpha_{3}\right\rangle$, $\left|\alpha_{3}\right\rangle\rightarrow \left|\alpha_{2}\right\rangle$, $\left|\alpha_{2}\right\rangle\rightarrow \left|\alpha_{1}\right\rangle$ and $\left|\alpha_{1}\right\rangle\rightarrow \left|\alpha_{0}\right\rangle$, this apperas as bit flip error. Since after four successive decay events the MPS $\left|\alpha_{j}\right\rangle$ recovers its initial state with decreased amplitude. For these reasons similar to the method described in ~\cite{33}, the bit flip error of MPS can be corrected by exploiting the effect of amplitude damping. On the other hand, when single photon based ququats under goes single decay event the decayed state can not be converted to initial ququat using unitary operations.

\section{Conclusion}
To conclude, we defined four new multi-photonic states with only 4$n+j$ (here, $j =0, 1, 2, 3$ and $n = 0, 1, ..., \infty$) number of photons. We showed how these multi-photonic states and another kind of entangled coherent states called bipartite four-component entangled coherent states can be generated using even superposition of coherent states as input to an interferometer. We also showed that the superposition of coherent states and bipartite four-component entangled coherent states can represent a ququat and an entangled ququat, respectively, using multi-photonic states as basis states of a four dimensional Hilbert space. We proposed a linear optical scheme for QT of such ququat using four-component entangled coherent states as quantum channel. Our scheme gives almost perfect teleportation with almost perfect success for appreciable coherent amplitude. We showed how such ququats encoded in superposition of coherent states can be used for realization of higher dimensional BB84 protocol in order to increase the security of distributed key. We also discussed the possible advantages of using ququats and entangled ququats based on superposition of coherent states in different quantum information processing tasks. These advantages arise due to the fact that entangled ququats based on superposition of coherent state defined in this paper can be completely discriminated by linear optical elements and photon detection. It has been proven earlier that by encoding information in higher dimensional quantum system like ququat, security of quantum communication can be increased and they also presents advantage in researches on foundations of quantum mechanics, particularly in study related to local realism. Theoretical study in this paper shows that superposition of coherent states may be used as a promising candidates for physical realization of ququat and entangled ququat, which in turn enable the encoding, processing and transmission of information more efficiently. 

\section{Acknowledgment}
We are grateful to Prof. N. Chandra and Prof. R. Prakash for their interest and stimulating discussions. The authors (MKM and VBJ) are also thankful to Dr. Raj Kumar, Dr. Arundhati Misra, Dr. A S Rajawat, Dr. R P Singh and Dr. P. K. Gupta for their support and encouragement.

\newpage
 \appendix 
 \section{}
 \label{}
\begin{spacing}{1}
\begin{table*}[htp]
\caption{Photon counting result 0 represent zero counts, and results 1, 2, 3 and 
4 represents nonzero results (1 modulo 4), (2 modulo 4), (3 modulo 4) and (0 
modulo 4), respectively. $B^{(j,k,m)}$ stands for Bob's state in mode 3 
after PC. $U^{(j,k)}$or $U^{(j,k,m)}$, stands for unitary operation to be 
performed by Bob to achieve teleported state $\left| T \right\rangle $ (defined 
in eq.~(\ref{eq25}))  with as large fidelity $F$ (defined in eq.~(\ref{eq26}))
as possible.}
\begin{center}
\begin{tabular}{|p{90pt}|p{48pt}|p{48pt}|p{48pt}|p{48pt}|}
\hline

Photon counting result& 

$B^{(j,k,m)}$& 

$U^{(j,k)}$or $U^{(j,k,m)}$& 

$\left| T \right\rangle $& 

$F$ \\

\hline

(4,4,0,0) (2,2,0,0) & 

$B^{(1,0,0)}$& 

$U^{(1,0)}$& 

\raisebox{-3ex}[0cm][0cm]{$\begin{array}{l}

 \left| {T_1 } \right\rangle  \\ 

 \end{array}$}& 

\raisebox{-3ex}[0cm][0cm]{$F_1$} \\

\cline{1-3} 

(0,0,4,4) (0,0,2,2)& 

$B^{(2,0,0)}$& 

$U^{(2,0)}$& 

 & 

  \\

\hline

(1,3,0,0) (3,1,0,0)& 

$-B^{(1,0,2)}$& 

$-U^{(1,0)}$& 

\raisebox{-3ex}[0cm][0cm]{$\begin{array}{l}

 \left| {T_2 } \right\rangle  \\ 

 \end{array}$}& 

\raisebox{-3ex}[0cm][0cm]{$F_2 $} \\

\cline{1-3} 

(0,0,1,3) (0,0,3,1)& 

$-B^{(2,0,2)}$& 

$-U^{(2,0)}$& 

 & 

  \\

\hline

(4,2,0,0) (2,4,0,0)& 

$-B^{(1,2,2)}$& 

$-U^{(1,2)}$& 

\raisebox{-3ex}[0cm][0cm]{$\begin{array}{l}

 \left| {T_3 } \right\rangle \\ 

 \end{array}$}& 

\raisebox{-3ex}[0cm][0cm]{$\begin{array}{l}

 F_3  \\ 

 \end{array}$} \\

\cline{1-3} 

(0,0,4,2) (0,0,2,4)& 

$-B^{(2,2,2)}$& 

$-U^{(2,2)}$& 

 & 

  \\

\hline

(1,1,0,0) (3,3,0,0)& 

$B^{(1,2,0)}$& 

$U^{(1,2)}$& 

\raisebox{-3ex}[0cm][0cm]{$\begin{array}{l}

 \left| {T_4 } \right\rangle \\ 

 \end{array}$}& 

\raisebox{-3ex}[0cm][0cm]{$\begin{array}{l}

 F_4  \\ 

 \end{array}$} \\

\cline{1-3} 

(0,0,1,1) (0,0,3,3)& 

$B^{(2,2,0)}$& 

$U^{(2,2)}$& 

 & 

  \\

\hline

(4,1,0,0) (2,3,0,0)& 

$\sqrt 2 B^{(3,1,3)}$& 

$\sqrt 2 U^{(3,1,3)}$& 

\raisebox{-8ex}[0cm][0cm]{$\begin{array}{l}

 \left| {T_5 } \right\rangle \\ 

 \end{array}$}& 

\raisebox{-8ex}[0cm][0cm]{$\begin{array}{l}

 F_5 \\ 

 \end{array}$} \\

\cline{1-3} 

(0,0,4,1) (0,0,2,3)& 

$\sqrt 2 B^{(0,1,1)}$& 

$\sqrt 2 U^{(0,1,1)}$& 

 & 

  \\

\cline{1-3} 

(1,4,0,0) (3,2,0,0)& 

$\sqrt 2 B^{(3,1,1)}$& 

$\sqrt 2 U^{(3,1,1)}$& 

 & 

  \\

\cline{1-3} 

(0,0,1,4) (0,0,3,2)& 

$\sqrt 2 B^{(0,1,3)}$& 

$\sqrt 2 U^{(0,1,3)}$& 

 & 

  \\

\hline

(4,3,0,0) (2,1,0,0)& 

$\sqrt 2 B^{(3,3,1)}$& 

$\sqrt 2 U^{(3,3,1)}$& 

\raisebox{-8ex}[0cm][0cm]{$\begin{array}{l}

 \left| {T_6 } \right\rangle \\ 

 \end{array}$}& 

\raisebox{-8ex}[0cm][0cm]{$\begin{array}{l}

 F_6 \\ 

 \end{array}$} \\

\cline{1-3} 

(0,0,4,3) (0,0,2,1)& 

$\sqrt 2 B^{(0,3,3)}$& 

$\sqrt 2 U^{(0,3,3)}$& 

 & 

  \\

\cline{1-3} 

(1,2,0,0) (3,4,0,0)& 

$\sqrt 2 B^{(3,3,3)}$& 

$\sqrt 2 U^{(3,3,3)}$& 

 & 

  \\

\cline{1-3} 

(0,0,1,2) (0,0,3,4)& 

$\sqrt 2 B^{(0,3,1)}$& 

$\sqrt 2 U^{(0,3,1)}$& 

 & 

  \\

\hline
\end{tabular}
\label{tab1}
\end{center}
\end{table*}
\end{spacing}
\begin{table*}[htp]
\caption{Photon counting result 0 represents zero photon counts, and results 1, 2, 3 and 
4 represents nonzero results (1 modulo 4), (2 modulo 4), (3 modulo 4) and (0 
modulo 4), respectively. $B^{(j,k)}$ stands for the Bob's state in mode 3 
after photon counting. $U^{(j,k)}$ stands for unitary operation to be performed by Bob to 
achieve teleported state with as large fidelity as possible.}
\begin{center}

\begin{tabular}{|p{190pt}|p{50pt}|p{56pt}|}

\hline

\textbf{Photon counting result in modes 8, 9, 10 {\&} 11}& 

$B^{(j,k)}$ & 

$U^{(j,k)}$  \\

\hline

(0,4,4,4) (0,4,2,2) (0,2,1,1) (0,2,3,3)& 

$B^{(2,0)}$& 

$U^{(2,0)}$ \\

\hline

(0,4,3,1) (0,4,1,3) (0,2,2,4) (0,2,4,2)& 

$-B^{(2,0)}$& 

$-U^{(2,0)}$ \\

\hline

(0,1,3,4) (0,1,1,2) (0,3,4,1) (0,3,2,3)& 

$iB^{(2,0)}$& 

$-iU^{(2,0)}$ \\

\hline

(0,1,2,1) (0,1,4,3) (0,3,1,4) (0,3,3,2)& 

$-iB^{(2,0)}$& 

$iU^{(2,0)}$ \\

\hline

(4,0,4,4) (4,0,3,1) (4,0,2,2) (4,0,4,3)& 

$B^{(0,0)}$& 

$U^{(0,0)}$ \\

\hline

(1,0,3,4) (1,0,2,1) (1,0,1,2) (1,0,4,3)& 

$iB^{(0,0)}$& 

$-iU^{(0,0)}$ \\

\hline

(2,0,2,4) (2,0,1,1) (2,0,4,2) (2,0,3,3)& 

$-B^{(0,0)}$& 

$-U^{(0,0)}$ \\

\hline

(3,0,1,4) (3,0,4,1) (3,0,3,2) (3,0,2,3)& 

$-iB^{(0,0)}$& 

$iU^{(0,0)}$ \\

\hline

(2,2,0,4) (2,4,0,2) (3,4,0,1) (3,2,0,3)& 

$B^{(1,0)}$& 

$U^{(1,0)}$  \\

\hline

(4,4,0,4) (4,2,0,2) (1,2,0,1) (1,4,0,3)& 

$B^{(1,0)}$& 

$U^{(1,0)}$ \\

\hline

(2,1,0,1) (2,3,0,3) (3,1,0,4) (3,3,0,2)& 

$-B^{(1,0)}$& 

$-U^{(1,0)}$ \\

\hline

(4,3,0,1) (4,1,0,3) (1,3,0,4) (1,1,0,2)& 

$-B^{(1,0)}$& 

$-U^{(1,0)}$ \\

\hline

(4,4,4,0) (4,3,1,0) (4,2,2,0) (4,1,3,0)& 

$B^{(3,0)}$& 

$U^{(3,0)}$  \\

\hline

(2,2,4,0) (2,1,1,0) (2,4,2,0) (2,3,3,0)& 

$B^{(3,0)}$& 

$U^{(3,0)}$ \\

\hline

(1,3,4,0) (1,2,1,0) (1,1,2,0) (1,4,3,0)& 

$B^{(3,0)}$& 

$U^{(3,0)}$ \\

\hline

(3,1,4,0) (3,4,1,0) (3,3,2,0) (3,2,3,0)& 

$B^{(3,0)}$& 

$U^{(3,0)}$ \\

\hline

(0,1,4,4) (0,1,2,2) (0,3,1,1) (0,3,3,3)& 

$B^{(2,1)}$& 

$U^{(2,1)}$ \\

\hline

(0,1,3,1) (0,1,1,3) (0,3,2,4) (0,3,4,2)& 

$-B^{(2,1)}$& 

$-U^{(2,1)}$ \\

\hline

(0,4,1,4) (0,4,3,2) (0,2,2,1) (0,2,4,3)& 

$-iB^{(2,1)}$& 

$iU^{(2,1)}$ \\

\hline

(0,4,4,1) (0,4,2,3) (0,2,1,2) (0,2,3,4)& 

$iB^{(2,1)}$& 

$-iU^{(2,1)}$ \\

\hline

(4,0,1,4) (4,0,4,1) (4,0,3,2) (4,0,2,3)& 

$B^{(0,1)}$& 

$U^{(0,1)}$  \\

\hline

(3,0,2,4) (3,0,1,1) (3,0,4,2) (3,0,3,3)& 

$-iB^{(0,1)}$& 

$iU^{(0,1)}$ \\

\hline

(2,0,3,4) (2,0,2,1) (2,0,1,2) (2,0,4,3)& 

$-B^{(0,1)}$& 

$-U^{(0,1)}$ \\

\hline

(1,0,4,4) (1,0,3,1) (1,0,2,2) (1,0,1,3)& 

$iB^{(0,1)}$& 

$-iU^{(0,1)}$ \\

\hline

(2,3,0,4) (2,1,0,2) (3,1,0,1) (3,3,0,3)& 

$B^{(1,1)}$& 

$U^{(1,1)}$ \\

\hline

(4,1,0,4) (4,3,0,2) (1,3,0,1) (1,1,0,3)& 

$B^{(1,1)}$& 

$U^{(1,1)}$ \\

\hline

(2,2,0,1) (2,4,0,3) (3,2,0,4) (3,4,0,2)& 

$-B^{(1,1)}$& 

$-U^{(1,1)}$ \\

\hline

(4,4,0,1) (4,2,0,3) (1,4,0,4) (1,2,0,2)& 

$-B^{(1,1)}$& 

$-U^{(1,1)}$ \\

\hline

(4,1,4,0) (4,4,1,0) (4,3,2,0) (4,2,3,0)& 

$B^{(3,1)}$& 

$U^{(3,1)}$ \\

\hline

(2,3,4,0) (2,2,1,0) (2,1,2,0) (2,4,3,0)& 

$B^{(3,1)}$& 

$U^{(3,1)}$ \\

\hline

\end{tabular}
\label{tab2}
\end{center}
\end{table*}
\begin{table*}[htp]
\caption{Table A.2 continude......}
\begin{center}
\begin{tabular}{|p{190pt}|p{50pt}|p{56pt}|}

\hline

\textbf{PC result in modes 8, 9, 10 {\&} 11}& 

$B^{(j,k)}$ & 

$U^{(j,k)}$  \\
\hline
(1,4,4,0) (1,3,1,0) (1,2,2,0) (1,1,3,0)& 

$B^{(3,1)}$& 

$U^{(3,1)}$ \\

\hline

(3,2,4,0) (3,1,1,0) (3,4,2,0) (3,3,3,0)& 

$B^{(3,1)}$& 

$U^{(3,1)}$ \\

\hline

(0,1,1,4) (0,1,3,2) (0,3,2,1) (0,3,4,3)& 

$B^{(2,2)}$& 

$U^{(2,2)}$ \\

\hline
(0,1,4,1) (0,1,2,3) (0,3,3,4) (0,3,1,2)& 

$-B^{(2,2)}$& 

$-U^{(2,2)}$ \\

\hline
(0,4,2,4) (0,4,4,2) (0,2,3,1) (0,2,1,3)& 

$-iB^{(2,2)}$& 

$iU^{(2,2)}$ \\

\hline
(0,4,4,1) (0,4,3,3) (0,2,4,4) (0,2,2,2)& 

$iB^{(2,2)}$& 

$-iU^{(2,2)}$ \\

\hline
(4,0,2,4) (4,0,1,1) (4,0,4,2) (4,0,3,3)& 

$B^{(0,2)}$& 

$U^{(0,2)}$ \\

\hline
(1,0,1,4) (1,0,4,1) (1,0,3,2) (1,0,2,3)& 

$iB^{(0,2)}$& 

$-iU^{(0,2)}$ \\

\hline
(2,0,4,4) (2,0,3,1) (2,0,2,2) (2,0,1,3)& 

$-B^{(0,2)}$& 

$-U^{(0,2)}$ \\

\hline
(3,0,3,4) (3,0,2,1) (3,0,1,2) (3,0,4,3)& 

$-iB^{(0,2)}$& 

$iU^{(0,2)}$ \\

\hline
(2,4,0,4) (2,2,0,2) (3,2,0,1) (3,4,0,3)& 

$B^{(1,2)}$& 

$U^{(1,2)}$ \\

\hline
(4,2,0,4) (4,4,0,2) (1,4,0,1) (1,2,0,3)& 

$B^{(1,2)}$& 

$U^{(1,2)}$ \\

\hline

(2,3,0,1) (4,4,0,2) (3,3,0,4) (3,1,0,2)& 

$-B^{(1,2)}$& 

$-U^{(1,2)}$ \\

\hline
(4,1,0,1) (2,1,0,3) (1,1,0,4) (1,3,0,2)& 

$-B^{(1,2)}$& 

$-U^{(1,2)}$ \\

\hline
(4,2,4,0) (4,1,1,0) (4,4,2,0) (4,3,3,0)& 

$B^{(3,2)}$& 

$U^{(3,2)}$ \\

\hline
(2,4,4,0) (2,3,1,0) (2,2,2,0) (2,1,3,0)& 

$B^{(3,2)}$& 

$U^{(3,2)}$ \\

\hline
(1,1,4,0) (1,4,1,0) (1,3,2,0) (1,2,3,0)& 

$B^{(3,2)}$& 

$U^{(3,2)}$ \\

\hline
(3,3,4,0) (3,2,1,0) (3,1,2,0) (3,4,3,0)& 

$B^{(3,2)}$& 

$U^{(3,2)}$ \\

\hline
(0,1,2,4) (0,1,4,2) (0,3,3,1) (0,3,1,3)& 

$B^{(2,3)}$& 

$U^{(2,3)}$ \\

\hline

(0,1,1,1) (0,1,3,3) (0,3,4,4) (0,3,2,2)& 

$-B^{(2,3)}$& 

$-U^{(2,3)}$ \\

\hline

(0,4,3,4) (0,4,1,2) (0,2,4,1) (0,2,2,3)& 

$-iB^{(2,3)}$& 

$iU^{(2,3)}$ \\

\hline

(0,4,2,1) (0,4,4,3) (0,2,1,4) (0,2,3,2)& 

$iB^{(2,3)}$& 

$-iU^{(2,3)}$ \\

\hline

(4,0,3,4) (4,0,2,1) (4,0,1,2) (4,0,4,3)& 

$B^{(0,3)}$& 

$U^{(0,3)}$ \\

\hline

(1,0,2,4) (1,0,1,1) (1,0,4,2) (1,0,3,3)& 

$iB^{(0,3)}$& 

$-iU^{(0,3)}$ \\

\hline
(2,0,1,4) (2,0,4,1) (2,0,3,2) (2,0,2,3)& 

$-B^{(0,3)}$& 

$-U^{(0,3)}$ \\

\hline

(3,0,4,4) (3,0,3,1) (3,0,2,2) (3,0,1,3)& 

$-iB^{(0,3)}$& 

$iU^{(0,3)}$ \\

\hline

(2,1,0,4) (2,3,0,2) (3,3,0,1) (3,1,0,3)& 

$B^{(1,3)}$& 

$U^{(1,3)}$ \\

\hline

(4,3,0,4) (4,1,0,2) (1,1,0,1) (1,3,0,3)& 

$B^{(1,3)}$& 

$U^{(1,3)}$ \\

\hline

(2,4,0,1) (2,2,0,2) (3,4,0,4) (3,2,0,2)& 

$-B^{(1,3)}$& 

$-U^{(1,3)}$ \\

\hline

(4,2,0,1) (4,4,0,3) (1,2,0,4) (1,4,0,2)& 

$-B^{(1,3)}$& 

$-U^{(1,3)}$ \\

\hline

(4,3,4,0) (4,2,1,0) (4,1,2,0) (4,4,3,0)& 

$B^{(3,3)}$& 

$U^{(3,3)}$ \\

\hline

(2,1,4,0) (2,4,1,0) (2,3,2,0) (2,2,3,0)& 

$B^{(3,3)}$& 

$U^{(3,3)}$ \\

\hline

(1,2,4,0) (1,1,1,0) (1,4,2,0) (1,3,3,0)& 

$B^{(3,3)}$& 

$U^{(3,3)}$ \\

\hline

(3,4,4,0) (3,3,1,0) (3,2,2,0) (3,1,3,0)& 

$B^{(3,3)}$& 

$U^{(3,3)}$ \\

\hline
\end{tabular}
\label{tab3}
\end{center}
\end{table*}

\end{document}